\documentclass[prc,floatfix,groupedaddress,nofootinbib,showpacs,preprintnumbers,
amsmath,amssymb,amsfonts,superscriptaddress] {revtex4-1}
\usepackage{graphicx}% Include figure files
\usepackage{dcolumn}%Align table columns on decimal point
\usepackage{mathrsfs}
\usepackage{bm}% bold math
\usepackage{graphicx}% Include figure files
\usepackage{dcolumn}% Align table columns on decimal point
\usepackage{bm}% bold math
\usepackage[usenames]{color}
\usepackage{relsize}

\def\expo{\mathlarger{\mathlarger{e}}}
\def\fF{\mathlarger{f}_{\raisebox{-0.5pt}{\!\tiny F}}}
\def\fS{\mathlarger{f}_{\raisebox{-0.5pt}{\!\tiny SF}}}
\newcommand{\qX}[1]{\mathlarger{q}_{\raisebox{-1.50pt}{\tiny #1}}}
\newcommand{\FF}[1]{\mathlarger{F}_{\raisebox{-0.5pt}{\!\tiny #1}}}
\newcommand{\rhoX}[1]{\mathlarger{\rho}_{\raisebox{-3.50pt}{\!\!\tiny #1}}}

\begin{document}

\title{The power of two: Assessing the impact of a second \\ measurement of the 
weak-charge form factor of ${}^{208}$Pb}
\author{J. Piekarewicz}
\email{jpiekarewicz@fsu.edu}
\affiliation{Department of Physics, Florida State University, Tallahassee, FL 32306}
\author{A. Linero}
\email{arlinero@stat.fsu.edu}
\affiliation{Department of Statistics, Florida State University, Tallahassee, FL 32306}
\author{P. Giuliani}
\email{pgg15@my.fsu.edu}
\affiliation{Department of Physics, Florida State University, Tallahassee, FL 32306}
\author{E. Chicken}
\email{chicken@stat.fsu.edu}
\affiliation{Department of Statistics, Florida State University, Tallahassee, FL 32306}

\date{\today}
\begin{abstract}
\begin{description}
\item[Background] Besides its intrinsic value as a fundamental nuclear-structure
	observable, the weak-charge density of ${}^{208}$Pb---a quantity that is 
	closely related to its neutron distribution---is of fundamental importance in
	constraining the equation of state of neutron-rich matter. 
\item[Purpose] To assess the impact that a second electroweak measurement 
	of the weak-charge form factor of ${}^{208}$Pb may have on the determination 
	of  its overall weak-charge density.
\item[Methods] Using the two putative experimental values of the form factor,
	together with a simple implementation of Bayes' theorem, we calibrate a 
	theoretically sound---yet surprisingly little known---\emph{symmetrized Fermi 
	function}, that is characterized by a density and form factor that are both 
	known exactly in closed form. 
\item[Results] Using the \emph{charge} form factor of ${}^{208}$Pb as a proxy
	for its weak-charge form factor, we demonstrate that using only two experimental 
	points to calibrate the symmetrized Fermi function is sufficient to accurately 
	reproduce the experimental charge form factor over a significant range of 
	momentum transfers.
\item[Conclusions] It is demonstrated that a second measurement of the
	weak-charge form factor of ${}^{208}$Pb supplemented by a robust 
	theoretical input in the form of the symmetrized Fermi function, would 
	place significant constraints on the neutron distribution of ${}^{208}$Pb
	and, ultimately, on the equation of state of neutron-rich matter.
\end{description}
\end{abstract}
%02.50.-r   %Probability theory, stochastic processes, and statistics
%02.60.Pn %Numerical optimization
%21.10.Ft	 Charge distribution
%21.10.Gv %nucleon distributions
%21.60.Jz %Nuclear Density Functional Theory
%21.65.-f  %Nuclear matter
%21.65.Cd %Asymmetric matter, neutron matter
% 21.65.Mn % Equations of state of nuclear matter
%21.65.Ef %Symmetry energy
%24.10.Jv %Relativistic models
%24.30.Cz %Giant resonances
%24.80.+y %nuclear tests of fundamental interactions and symmetries
%25.30.Bf %Elastic electron scattering
%26.60.Gj %Neutron-star crust
%26.60.Dd %Neutron-star core
%26.60.Kp %Neutron-star EOS
%26.60.-c  %Nuclear matter aspects of Neutron stars
%97.60.Jd %Neutron stars,
\maketitle

%%%%%  SECTION 1 - Introduction  %%%%%
\section{Introduction}
\label{intro}

Starting with the pioneering work of Hofstadter in the late 1950's \cite{Hofstadter:1956qs} 
and continuing until this day\,\cite{DeJager:1987qc,Fricke:1995,Angeli:2013}, elastic 
electron scattering has painted the most accurate and detailed picture of the distribution 
of protons in the atomic nucleus. This sits in stark contrast to our poor knowledge of the
neutron distribution which until very recently has been mapped using exclusively hadronic 
experiments that are hindered by large and uncontrolled uncertainties\,\cite{Horowitz:2014bja}. 
The Lead Radius EXperiment (``PREX'') at the Jefferson Laboratory opened a new window
by using parity-violating elastic electron scattering to provide the first model-independent 
determination of the weak form factor of ${}^{208}$Pb, albeit at a single value of the 
momentum transfer\,\cite{Abrahamyan:2012gp,Horowitz:2012tj}. 
Given that the weak charge of the neutron is much larger than the corresponding one 
of the proton, parity-violating electron scattering provides an ideal \emph{electroweak} 
probe of the neutron distribution\,\cite{Donnelly:1989qs}. Although measuring the weak 
charge form factor at a single point provides limited information on the neutron distribution, 
by invoking some theoretical assumptions, PREX furnished the first credible 
estimate of the neutron radius of ${}^{208}$Pb ($R_{n}^{208}$)\,\cite{Horowitz:2012tj}. Since
the proton radius of ${}^{208}$Pb ($R_{p}^{208}$) is known with enormous 
accuracy\,\cite{Angeli:2013}, PREX effectively determined the neutron skin thickness of 
${}^{208}$Pb\,\cite{Abrahamyan:2012gp,Horowitz:2012tj}:
%%%
\begin{equation}
 R_{\rm skin}^{208} \equiv R_{n}^{208}-R_{p}^{208}=
 {0.33}^{+0.16}_{-0.18}\,{\rm fm}.
\label{Rskin208}
\end{equation}
%%%

The determination of the neutron skin thickness of ${}^{208}$Pb is of great significance 
for multiple reasons. First, as an observable sensitive to the \emph{difference} between 
the neutron and proton densities, it plays a critical role in constraining the isovector sector 
of the nuclear energy density functional\,\cite{Reinhard:2010wz,Reinhard:2013fpa,
Nazarewicz:2013gda,Chen:2014sca,Chen:2014mza}. Second, a very strong correlation
has been found between the slope of the symmetry energy at saturation density ($L$)
and $R_{\rm skin}^{208}$\,\cite{Brown:2000, Furnstahl:2001un,Centelles:2008vu,
RocaMaza:2011pm,Chen:2014sca}. This provides a powerful connection between a
fundamental parameter of the equation of state (EOS) and a laboratory observable. 
Note that $L$ is closely related to the pressure of pure neutron matter at saturation 
density. Third, constraining the EOS of neutron-rich matter provides critical guidance on 
the interpretation of heavy-ion experiments involving nuclei with large neutron-proton
asymmetries\,\cite{Tsang:2004zz,Chen:2004si,Steiner:2005rd,Shetty:2007zg,Tsang:2008fd,
Li:2008gp}. Finally, even though there is a difference in length scales of 18 orders of 
magnitude, the neutron skin thickness of ${}^{208}$Pb and the radius of a neutron star
share a common dynamical origin\,\cite{Horowitz:2000xj,Horowitz:2001ya,
Carriere:2002bx,Steiner:2004fi,Li:2005sr,Erler:2012qd,Chen:2014sca}. Although in 
general neutron-star properties are sensitive to the high-density component of the EOS, 
it is the pressure in the neighborhood of twice nuclear matter saturation density that sets 
the overall scale for stellar radii\,\cite{Lattimer:2006xb}. Thus, whether pushing against 
surface tension in a nucleus or against gravity in a neutron star, it is the pressure in this 
neighborhood that determines both the thickness of the neutron skin and the radius of a 
neutron star.

However, the accurate and reliable determination of both the neutron skin thickness of
${}^{208}$Pb and the radius of a neutron star present enormous challenges. While PREX 
convincingly demonstrated the feasibility of the method for measuring weak-charge form
factors with an excellent control of systematic errors\,\cite{Abrahamyan:2012gp,Horowitz:2012tj}, 
unforeseen technical problems compromised the statistical accuracy of the experiment;
see the large errors in Eq.\,(\ref{Rskin208}). Fortunately, it is anticipated that in an already 
approved experiment (``PREX-II'') the uncertainty in the determination of $R_{n}^{208}$ 
will be reduced by a factor of three, to about $\pm\,0.06\,{\rm fm}$. In turn, attempts to reliably 
extract stellar radii have been hindered by large systematic uncertainties that have resulted 
in an enormous disparity---ranging from radii as small as 8\,km all the way to 
14\,km\,\cite{Ozel:2010fw,Steiner:2010fz,Suleimanov:2010th}. And whereas a better 
understanding of systematic uncertainties, new theoretical developments, and the 
implementation of robust statistical methods seem to favor small stellar 
radii\,\cite{Guillot:2013wu,Lattimer:2013hma,Heinke:2014xaa,Guillot:2014lla,
Ozel:2015fia}, a consensus has yet to be reached. Thankfully, the historical first 
detection of gravitational waves\,\cite{Abbott:PRL2016} opens a new window into 
the structure of neutron stars---particularly stellar radii---from the gravitational-wave
signal from the merger of two neutron stars; see 
Refs.\,\cite{Bauswein:2011tp,Lackey:2014fwa} and references contained therein.
Indeed, it is anticipated that gravitational waves could constrain neutron-star 
radii to better than one kilometer\,\cite{Lackey:2014fwa}.

As alluded to earlier, the pioneering PREX experiment measured the weak form 
factor of ${}^{208}$Pb at the single momentum transfer of 
$q\!=\!0.475\,{\rm fm}^{-1}$\,\cite{Abrahamyan:2012gp,Horowitz:2012tj}. The main 
goal of this contribution is to assess the impact that a second measurement could
have in determining the weak-form factor of ${}^{208}$Pb and, ultimately, in constraining 
the poorly-known density dependence of the symmetry energy. The need for 
a second measurement may be justified using simple arguments based on the 
nuclear mass formula of Bethe and Weizs\"acker\,\cite{Weizsacker:1935,Bethe:1936}. 
Given that the nuclear force saturates, Bethe and Weizs\"acker modeled the atomic 
nucleus as an incompressible liquid drop. The nearly uniform density found in the interior 
of a heavy nucleus features among the most successful predictions of the model. 
However, the liquid drop is finite so a penalty must be assessed for the formation 
of the nuclear surface. In this way, the density of a heavy nucleus is largely 
characterized by two parameters: a \emph{radius} that accounts for the distance 
over which the density is nearly uniform and a \emph{surface thickness} that controls 
the transition from high to low density. The corresponding nuclear form factor---which is 
the physical observable that is actually probed in the experiment---is obtained from the 
Fourier transform of the density distribution.  As such, the radius and the surface thickness 
leave a very distinct imprint on the form factor. Indeed, the nuclear form factor displays a 
nearly universal behavior characterized by diffractive oscillations controlled by the radius 
that are in turn modulated by an exponential falloff controlled by the surface 
thickness\,\cite{Amado:1986pm}.

Clearly, a single measurement of the form factor can only constrain a  linear 
combination of the radius and the surface thickness. This hinders the 
model-independent determination of the mean-square radius of the distribution 
as one must rely on theoretical models to lift the ``degeneracy". Instead, a second 
measurement of the form factor will allow the experimental determination of these 
two critical parameters. In this contribution we assess the impact of a second 
measurement of the weak-charge form factor of ${}^{208}$Pb by introducing, 
or rather re-introducing, a highly convenient two-parameter characterization of the 
density distribution: the \emph{symmetrized Fermi function}; see 
Ref.\,\cite{Sprung:1997} and references contained therein. For heavy nuclei
with a radius parameter that is significantly larger than the surface thickness,
the symmetrized Fermi (SFermi) function is practically indistinguishable from 
the standard Fermi (or Woods-Saxon) parametrization. However, unlike the 
standard Fermi function that displays a ``cusp'' at the center of the nucleus, the 
SFermi parametrization is analytic. In the present context, this offers a unique 
advantage over the standard Fermi function: the form factor associated to the 
symmetrized Fermi function can be computed exactly in closed form. That this 
elegant result remains largely unknown to the nuclear physics community comes 
as a surprise (although see Ref.\,\cite{Jiang:2010zzg}). As stated in 
Ref.\,\cite{Sprung:1997}: ``The symmetrized 
Fermi function has been known to some experts, but the least one can 
say is that it is not \emph{well known} generally. None of the text books 
on nuclear physics refers to it". On the other hand, a well-known parametrization 
of the nuclear form factor---with a density  that is also known in closed form---was 
introduced by Helm almost 6 decades ago\,\cite{Helm:1956zz}. However, the Helm 
form factor has a Gaussian falloff rather than the more realistic exponential falloff 
displayed by the SFermi form factor.

We have organized the paper as follows: In Sec.\,\ref{formal} we introduce the SFermi 
and Helm parametrization, and discuss the simple formalism used in the optimization of 
the two parameters that define these parametrizations. Sec.\,\ref{results} presents in a
simple, yet statistically rigorous manner, the great improvement in the determination 
of the form factor once a suitably chosen second measurement has been selected. 
To establish this point, we use the well known charge form factor of ${}^{208}$Pb as a 
proxy for its weak-charge form factor. Finally, we offer our summary and conclusions 
in Sec.\,\ref{conclusions}.

%%%%%  SECTION 2 - Formalism  %%%%%
\section{Theoretical Formalism}
\label{formal}

In this section we develop the formalism required to assess the role of a
second electroweak measurement of the weak form factor of ${}^{208}$Pb.
In the first part of the section we introduce the SFermi and Helm functions
underscoring that both have density distributions and form factors that are 
known in closed analytic form. The second part of this section discusses 
the simple Bayesian approach that we implement to constrain the two 
parameters that define the SFermi and Helm form factors.

\subsection{Symmetrized Fermi Form Factor}
\label{sfermi}

The Woods-Saxon, or Fermi, function was introduced more than six decades 
ago to describe nucleon-nucleus scattering\,\cite{Woods:1954zz}. Given that 
the nucleon-nucleon interaction is of short range relative to the overall size of 
the nucleus, the mean-field potential that the nucleon scatters from resembles 
the underlying nuclear density that is fairly accurately described in terms of a 
two-parameter Fermi shape. The conventional Fermi function is defined as
follows:
%%%
\begin{equation}
 \fF(r) = \frac{1}{1+\expo^{(r-c)/a}}\,, 
\end{equation}
%%%
where $c$ is the ``half-density radius" and $a$ the ``surface diffuseness''.  

A less known distribution that is practically identical to the Fermi function 
in the relevant nuclear domain of $c\!\gg\!a$ is the 
\emph{symmetrized Fermi} function:
%%%
\begin{equation}
 \fS(r) \equiv  \fF(r) + \fF(-r) - 1 =
 \left(\frac{1}{1+\expo^{(r-c)/a}} - \frac{1}{1+\expo^{(r+c)/a}}\right) =
 \frac{\sinh(c/a)}{\cosh{(r/a)}+\cosh(c/a)}\,.
\end{equation}
%%%

For an enlightening introduction to the SFermi distribution that underscores
its unique analytic behavior see Ref.\,\cite{Sprung:1997} and references 
contained therein. Although practically indistinguishable from the conventional 
Fermi function, the symmetrized Fermi function enjoys a distinct advantage 
over it: whereas the Fermi function displays a cusp at the origin, the derivative 
of the SFermi function vanishes smoothly at $r\!=\!0$. As a result of the analyticity 
of the SFermi function, its form factor---namely, the Fourier transform of the
one-body density---may be, unlike the case of the conventional Fermi function, 
\emph{evaluated in closed analytic form}\,\cite{Sprung:1997}. 
That is,
%%%
\begin{align}
 \FF{SF}(q) &= \int e^{-i{\bf q}\cdot{\bf r}} \rhoX{SF}(r) d^{3}r =
 4\pi\int_{0}^{\infty} \frac{\sin(qr)}{qr} \rhoX{SF}(r) r^{2} dr \nonumber\\
           & = \frac{3}{qc\Big((qc)^{2}+(\pi qa)^{2}\Big)}       
                  \left(\frac{\pi qa}{\sinh(\pi qa)}\right)
                  \left[\frac{\pi qa}{\tanh(\pi qa)}\sin(qc)-qc\cos(qc)\right]\,.
 \label{FFSF}                  
\end{align}
%%%
where
%%%
\begin{equation}
 \rhoX{SF}(r) \equiv \rhoX{0}\fS(r); \quad
 \rhoX{0}\equiv\frac{3}{4\pi c\left(c^{2}+\pi^{2}a^{2}\right)} \,,
 \label{RhoSF} 
\end{equation}
%%%
and we have adopted the following normalization: 
%%%
\begin{equation}
 \FF{SF}(q\!=\!0) = \int\!\rhoX{SF}(r) d^{3}r = 1\,.
 \label{Normal}
\end{equation}
%%%
Among the many appealing features of an analytic expression for the form 
factor is that all the moments of the distribution can be evaluated exactly. 
Indeed, for low momentum transfers a Taylor series expansion of the form 
factor yields
%%%
\begin{equation}
 \FF{SF}(q) = 1 - \frac{q^{2}}{3!}R^{2} + \frac{q^{4}}{5!}R^{4} -
 \frac{q^{6}}{7!}R^{6} + \ldots
 \label{LowqFs}
\end{equation}
%%%
where the first three moments of the SFermi distribution are given by
%%%
\begin{subequations}
\begin{align}
 R^{2}  & \equiv \langle r^{2} \rangle  = \frac{3}{5}c^{2} + 
  \frac{7}{5}(\pi a)^{2} \,,\\
 R^{4} & \equiv \langle r^{4} \rangle  = \frac{3}{7}c^{4} + 
  \frac{18}{7}(\pi a)^{2}c^{2} + \frac{31}{7}(\pi a)^{4}  \,,\\
 R^{6} & \equiv\langle r^{6} \rangle  = \frac{1}{3}c^{6} + 
  \frac{11}{3}(\pi a)^{2}c^{4} + \frac{239}{15}(\pi a)^{4}c^{2} + 
  \frac{127}{5}(\pi a)^{6} \,.
\end{align}
 \label{SFMoments}
\end{subequations}
%%%
Unlike the conventional Fermi function, these expressions---and indeed all the moments of 
the SFermi distribution---are exact as they do not rely on a power series expansion in terms 
of the ``small'' parameter $\pi a/c$. Also interesting and highly insightful is the behavior of 
the SFermi form factor in the limit of high momentum transfers. Indeed, in this limit the 
SFermi form factor takes a remarkably simple form
%%%
\begin{equation}
 \FF{SF}(q) \rightarrow
 -6\frac{\pi a}{\sqrt{c^{2}+\pi^{2}a^{2}}}\frac{\cos(qc+\delta)}{qc}\,
 \mathlarger{e^{-\pi qa}}\,; \quad 
 \tan\delta\!\equiv\!\frac{\pi a}{c}\,.
  \label{HighqFs}
\end{equation}
%%%
This expression encapsulates many of the insights developed more than 
three decades ago in the context of the conventional Fermi function. 
Namely,  that for large momentum transfers the oscillations in the form 
factor are controlled by the half-density radius $c$ and the exponential 
falloff by the diffuseness parameter $a$ (or rather 
$\pi a$)\,\cite{Amado:1979st,Amado:1986pm}. Again, it should be 
underscored that this expression is exact in the limit of high momentum 
transfers.

\subsection{Helm Form Factor}
\label{helm}

Another simple, yet realistic, distribution that also captures the main features 
of the form factor is the Helm function. Although much better known than the 
symmetrized Fermi form factor, in the interest of completeness we provide a 
short summary of its most important properties. The Helm form factor was 
introduced exactly 60 years ago to analyze elastic scattering of electrons from 
nuclei\,\cite{Helm:1956zz}. The Helm form factor is defined as the product of 
two fairly simple form factors: one associated with a uniform (``box'') density 
and the other one accounting for a \emph{Gaussian} 
falloff\,\cite{Helm:1956zz, Mizutori:1999ai,Horowitz:2012tj}. That is,
%%%
\begin{equation}
 \FF{H}(q) =\FF{B}(q)\FF{G}(q) = 
 3\,\frac{j_{1}(qR_{0})}{qR_{0}} \mathlarger{e^{-q^{2}\sigma^{2}/2}}\,,
 \label{Helm}
\end{equation}
%%%
where
%%%
\begin{subequations}
\begin{align}
 \FF{B}(q) & = \int e^{-i{\bf q}\cdot{\bf r}} \rhoX{B}(r) d^{3}r =
  \int e^{-i{\bf q}\cdot{\bf r}} 
 \left(\frac{3\Theta(R_{0}\!-\!r)}{4\pi R_{0}^{3}}\right) d^{3}r =
  3\,\frac{j_{1}(qR_{0})}{qR_{0}}\,,\\ 
 \FF{G}(q) & = \int e^{-i{\bf q}\cdot{\bf r}} \rhoX{G}(r) d^{3}r =
 \int e^{-i{\bf q}\cdot{\bf r}} 
 \left(\frac{e^{-r^{2}/2\sigma^{2}}}{(2\pi\sigma^{2})^{3/2}}\right) d^{3}r =
 \mathlarger{e^{-q^{2}\sigma^{2}/2}} \,.
  \label{HelmFFs}
\end{align}
\end{subequations}
%%%
Here $j_{1}(x)$ is the spherical Bessel function of order one:
%%%
\begin{equation}
 j_{1}(x) = \frac{\sin(x)}{x^{2}}-\frac{\cos(x)}{x} \,.
 \label{Jone}
\end{equation}
%%%
A great advantage of the Helm form factor is that it is defined 
in terms of a form factor that encodes the uniform interior density 
and another one that characterizes the nuclear surface. As such, 
the Helm form factor is defined entirely in terms of two constants: 
the box (or ``diffraction'') radius $R_{0}$ and the surface 
thickness $\sigma$. Although slightly more complicated than the 
form factor, a closed-form expression for the Helm density also 
exists. It is given by,
%%%
\begin{equation}
 \rhoX{H}(r) \!=\! \frac{1}{2}\rhoX{0}
 \left[{\rm erf}\!\left(\frac{r+R_{0}}{\sqrt{2}\sigma}\right) \!-\!
 {\rm erf}\!\left(\frac{r-R_{0}}{\sqrt{2}\sigma}\right)\right] \!+\!
 \frac{1}{\sqrt{2\pi}}\left(\frac{\sigma}{r}\right)\rhoX{0}
 \left[\exp\!\left(-\frac{(r+R_{0})^{2}}{2\sigma^{2}}\right) \!-\!
 \exp\!\left(-\frac{(r-R_{0})^{2}}{2\sigma^{2}}\right)\right]; 
 \hspace{3pt}
 \rhoX{0}\equiv\frac{3}{4\pi R_{0}^{3}},
 \label{HelmD}
\end{equation}
%%%
where ${\rm erf}(x)$ is the error function
%%%
\begin{equation}
 {\rm erf}(x) = \frac{2}{\sqrt{\pi}}\int_{0}^{x} 
 \mathlarger{e^{-z^{2}}} dz. 
 \label{ErrorF}
\end{equation}
%%%
As in the case of the symmetrized Fermi function, the Helm form 
factor has been normalized to $\FF{H}(q\!=\!0)\!=\!1$. Finally, the 
first three moments of the Helm distribution are given by the following 
simple expressions:
%%%
\begin{subequations}
\begin{align}
 R^{2}  & \equiv \langle r^{2} \rangle  = 
  \frac{3}{5}R_{0}^{2} + 3\sigma^{2} \,,\\
 R^{4} & \equiv \langle r^{4} \rangle  = \frac{3}{7}R_{0}^{4} + 
  6R_{0}^{2}\sigma^{2} + 15\sigma^{4}  \,,\\
 R^{6} & \equiv\langle r^{6} \rangle  = \frac{1}{3}R_{0}^{6} + 
  9R_{0}^{4}\sigma^{2} + 63R_{0}^{2}\sigma^{4} + 
  105\sigma^{6} \,.
\end{align}
\label{HelmMoments}
\end{subequations}

\subsection{Parameter Optimization}
\label{ParOpt}
 
In this section we outline the necessary steps that are required to determine 
the two model parameters that define the SFermi and Helm form factors from a 
measurement of the experimental weak-charge form factor of ${}^{208}$Pb. 
Evidently, without further theoretical assumptions, it is impossible to constrain 
both model parameters from our current knowledge, namely, a single 
measurement of the form factor; see Eqs.\,(\ref{SFMoments}) 
and\,(\ref{HelmMoments}). In the particular case of PREX---where the weak 
form factor was extracted at a single $q$-point---constraints on the surface 
thickness $\sigma$ of the Helm model were obtained by analyzing the 
theoretical predictions of several mean-field models. This lead to a 
\emph{theoretical} uncertainty in the determination of $\sigma$ of about 
10\%\,\cite{Horowitz:2012tj}, which was ultimately incorporated into the final 
estimate of the weak-charge radius of ${}^{208}$Pb. Our aim here is to 
demonstrate that measuring the weak form factor at a suitable second  
point minimizes the reliance on theoretical models.

Naturally, the selection of the first $q$-point should match the PREX 
momentum transfer of $\qX{1}\!=\!0.475\,{\rm fm}^{-1}$. Given this 
unique data point, how accurately can we constrain the weak form 
factor of ${}^{208}$Pb and in particular its mean-square radius? To
answer this question from a strict statistical perspective we must 
construct the \emph{likelihood function} defined as\,\cite{Stone:2013}: 
%%%
\begin{equation}
 p(F|\omega) = \mathlarger{e^{-\frac{1}{2}\chi^2(F;\,\omega)}} \;,
 \label{Likelihood}
\end{equation}
%%%
where
%%%
\begin{equation}
 \chi^{2}(F;\omega) = 
 \frac{\Big(\FF{SF}(\qX{1}; \omega) - F_{\rm exp}(\qX{1})\Big)^{2}}
 {\Delta F_{\rm exp}^{2}(\qX{1})} \,.
 \label{ChiSquareOnePoint}
\end{equation}
%%%
Here $\Delta F_{\rm exp}$ defines the experimental error and
$\omega\!\equiv\!\{a,c\}$ denotes the two model parameters of the 
symmetrized Fermi function $\FF{SF}(q)$. Note that the likelihood 
function $p(F|\omega)$ represents the probability 
density that a given set of parameters $\omega$ reproduces the 
experimental form factor $F_{\rm exp}(q)$ at the given PREX point.
Often, however, one may refine the probability distribution by injecting 
our own biases and intuition. For example, as indicated in 
Ref.\,\cite{Horowitz:2012tj}, mean-field predictions of the weak-charge 
density of ${}^{208}$Pb suggest a Helm surface thickness of  
$\sigma\!=\!(1.02\pm0.09)$\,fm. Such biases may then be incorporated 
into a \emph{prior} probability $p(\omega)$ that represents the best 
estimate of the model parameters prior to the realization of the 
experiment(s). For the prior we adopt a fairly broad Gaussian 
distribution (see Table\,\ref{Table1}) centered around the predictions 
of an accurately-calibrated set of mean-field models\,\cite{Chen:2014mza}. 
That is,
%%%
\begin{equation}
 p(\omega) = \prod_{i} \frac{1}{\sqrt{2\pi\sigma_{i}^{2}}}\,
 \mathlarger{e^{-(\omega_{i}-\bar{\omega}_{i})^{2}/2\sigma_{i}^{2}}} \;.
 \label{Prior}
\end{equation}
%%%
To provide the connection between our own theoretical biases (encoded
in the prior) and the experimental measurement (encoded in the likelihood) 
we invoke Bayes' theorem. That is,
%%%
\begin{equation}
 p(\omega|F)=\frac{p(F|\omega)\hspace{1pt}p(\omega)}{p(F)},
 \label{Posterior}
\end{equation} 
%%%
where $p(F)$ is known as the \emph{marginal likelihood}\,\cite{Stone:2013}. 
The \emph{posterior} probability density $p(\omega|F)$, namely the 
improvement in our prior knowledge as a result of the measurement, represents 
the probability density that a given set of SFermi parameters describes the measured
experimental form factor. In principle, the implementation of Bayes' theorem 
requires the specification of the marginal likelihood. In practice, however,
this term---as well as any other normalization factor independent of 
$\omega$---may be disregarded as long as we are only interested in the
relative probabilities that will be computed using Monte Carlo methods.

As we shall see later in Sec.\,\ref{results}, the posterior probability density 
defined in such a way provides a robust statistical benchmark for the selection 
of the second $q$-point\,\cite{Chaloner:1995}. Our goal in selecting this second 
point is to lift the degeneracy among the many models that are consistent with 
the single PREX measurement. To do so, one should search for a region in $q$ 
that maximizes the variability among the theoretical predictions. Perhaps not 
surprising, this happens near the first diffraction maximum ({\sl i.e.,} near 
the first maximum in $|F(q)|$ away from $q\!=\!0$). Once the two values 
of the momentum transfer ($\qX{1}$ and $\qX{2}$) have been selected, 
the determination of the model parameters follows from a likelihood 
function suitably augmented relative to 
Eq.\,(\ref{ChiSquareOnePoint}). That is, the augmented objective 
function $\chi^{2}(F;\omega)$ now becomes
%%%
\begin{equation}
 \chi^{2}(F;\omega) = 
 \frac{\Big(\FF{SF}(\qX{1}; \omega) - F_{\rm exp}(\qX{1})\Big)^{2}}
 {\Delta F_{\rm exp}^{2}(\qX{1})} +
 \frac{\Big(\FF{SF}(\qX{2}; \omega) - F_{\rm exp}(\qX{2})\Big)^{2}}
 {\Delta F_{\rm exp}^{2}(\qX{2})} \,.
 \label{ChiSquareTwoPoints}
\end{equation}
%%%

%%%%%  SECTION 3 - Results  %%%%%
\section{Results}
\label{results}

In this section we examine the accuracy that may be attained in the determination 
of the weak form factor of ${}^{208}$Pb and more specifically in its weak-charge 
radius from only two experimental measurements. To test the soundness and 
reliability of the proposed method, we rely on a form factor that is known with 
exquisite precision over many orders of magnitude in momentum transfer: the 
\emph{charge} form factor of ${}^{208}$Pb\,\cite{DeJager:1987qc}. That is, we 
simulate the impact of a second measurement by selecting the charge form factor 
of ${}^{208}$Pb as a proxy for the weak form factor. 
%%%%%%%%%%%%%%%%%%%%%%%%%%%%%%%%%%%%%%%%%%%%%%%%%%%%%
\subsection{Selection of the second value of the momentum transfer}
\label{q2selection}

The selection of the second value of the momentum transfer $\qX{2}$ is 
motivated by our desire to lift the degeneracy between the many models
that satisfy the constraint imposed by the single PREX measurement. To
do so, we search in a region of $q$ that maximizes the variability among 
the theoretical predictions. Namely, we define the one-point likelihood in
Eq.\,(\ref{ChiSquareOnePoint}) in terms of the charge form factor of ${}^{208}$Pb
at $\qX{1}\!=\!0.5\,{\rm fm}^{-1}$ which is given by $\FF{ch}(\qX{1})\!=\!0.210$. 
This choice is motivated by the existing PREX\,\cite{Abrahamyan:2012gp} 
measurement that resulted in a weak form factor comparable to $\FF{ch}(\qX{1})$, 
namely,
$\FF{wk}(\qX{1})\!=\!0.204\pm0.028\,({\rm exp})\pm(0.001)\,({\rm model})$\,\cite{Horowitz:2012tj}. 
For the experimental uncertainty we assume $\Delta F_{\rm exp}(\qX{1})\!=\!0.005$. 
Although significantly larger than the typical error of the charge form factor, attaining 
such level of precision may be difficult for the weak-charge form factor---at least 
as presently envisioned by PREX-II. However, such precision may be achieved at 
the planned ``Mainz Energy-Recovering Superconducting Accelerator'' (MESA) 
facility. Finally, we adopt what we view as a fairly conservative choice for the 
Gaussian prior $p(\omega)$ defined in Eq.\,(\ref{Prior}). Whereas for the central 
values $\bar{\omega}_{i}$ we rely on the predictions of an accurately calibrated 
model\,\cite{Chen:2014mza}, for the \emph{relative} standard deviation we adopt 
a fixed value of one half to account for the significant spread displayed by a large 
collections of mean-field models\,\cite{Centelles:2008vu,RocaMaza:2011pm}. 
The (prior) central values for both the SFermi and Helm distributions have been 
tabulated in Table\,\ref{Table1}. We note that we have tested other choices and 
found our results robust against the choice of prior.

%%%%%%%%%%%%%%%%%%%%%%%%%%%%%%%%%%%
\begin{table}[h]
  \begin{tabular}{|r|r|}
   \hline
    SFermi\hspace{6pt} & Helm\hspace{9pt}  \\
   \hline
    \rule{0pt}{9pt}  
      $\bar{c}=6.683$ &  $\bar{R}_{0}=6.788$    \\
      $\bar{a}=0.494$ &  $\bar{\sigma}=0.880$   \\
   \hline
  \end{tabular}
  \caption{Central values for the Gaussian distribution assumed for the prior;
  see Eq.\,(\ref{Prior}). In all cases the \emph{relative} standard deviation has
  been fixed to one half; for example, $\sigma_{\!c}\!=\!\bar{c}/2\!=\!3.342\,{\rm fm}$. 
  All quantities are given in fm.}
  \label{Table1}
 \end{table}
%%%%%%%%%%%%%%%%%%%%%%%%%%%%%%%%%%%

%%%%%%%%%%%%%%%%%%%%%%%%%%%%%%%%%%%%%%%%%%%%%%%%%%%%%
\begin{figure}[ht]
 \begin{center}
  \includegraphics[width=0.45\linewidth]{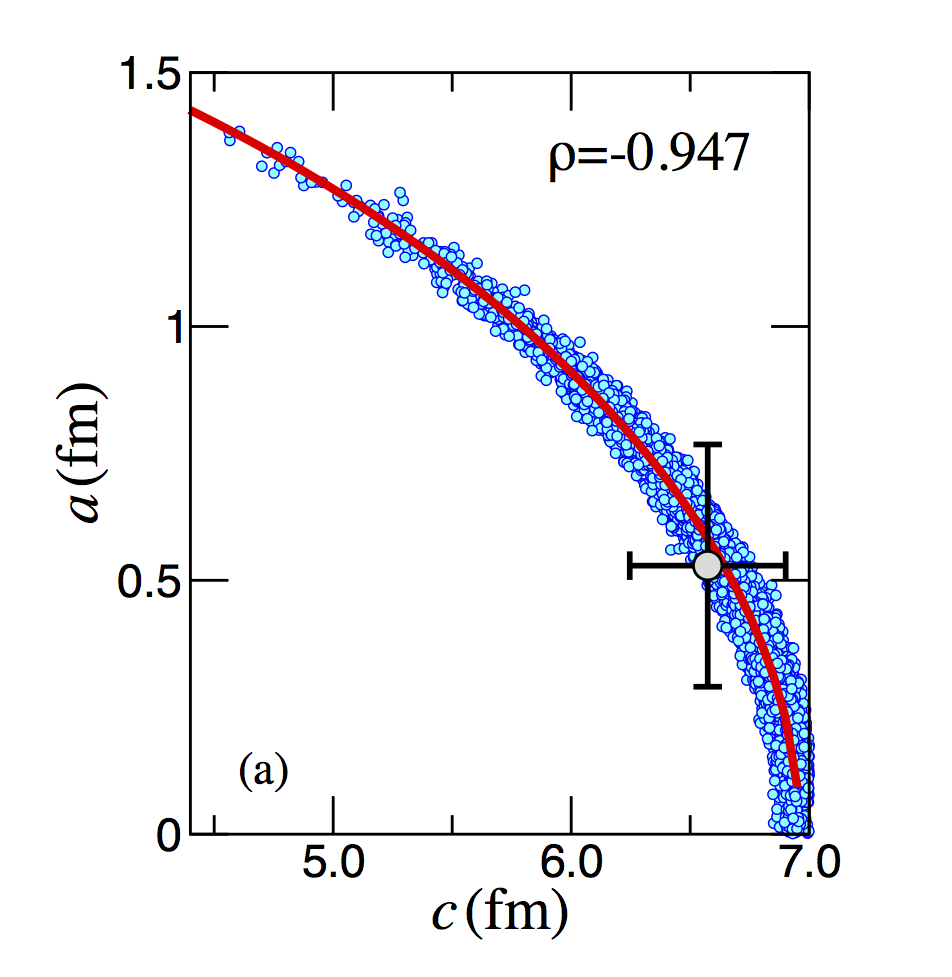} 
  \includegraphics[width=0.45\linewidth]{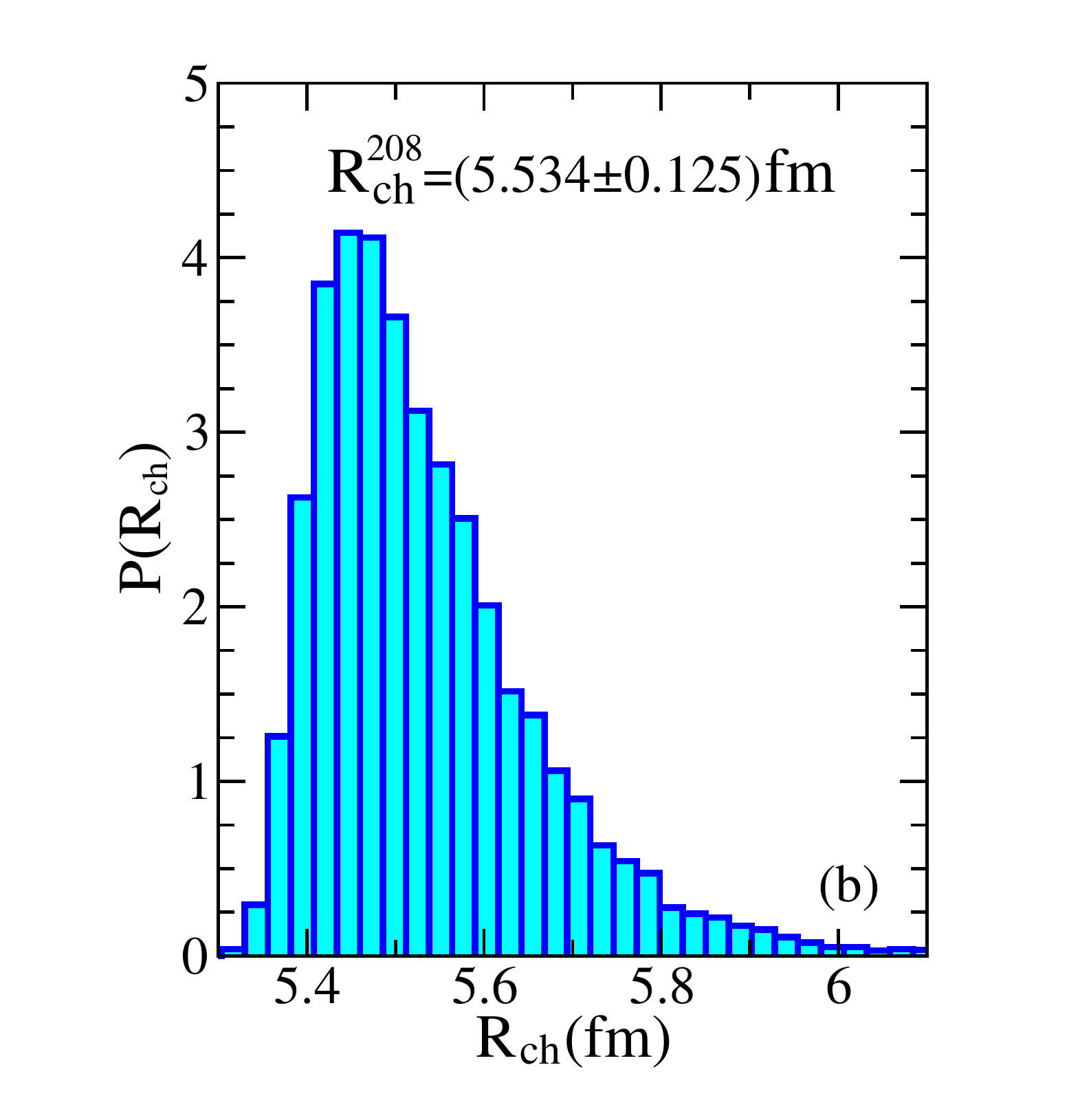}
  \caption{(color online) (a) Correlation plot between the half-density radius $c$ and the 
  surface diffuseness $a$ that define the symmetrized Fermi function. The number
  of points represent the raw data obtained from the Monte-Carlo simulation. Also shown 
  are the respective averages and standard deviations. The solid red line represents the 
  functional form obtained from solving the equation $F_{\rm ch}(q_{1};a,c)\!=\!0.210$. (b) 
  Probability distribution function for the charge radius of ${}^{208}$Pb obtained from the 
  Monte-Carlo simulation.}
  \label{Fig1}
 \end{center} 
\end{figure}
%%%%%%%%%%%%%%%%%%%%%%%%%%%%%%%%%%%%%%%%%%%%%%%%%%%%%

With the definition of the posterior distribution now firmly in place, we 
proceed to generate the distribution of SFermi parameters using a standard 
Metropolis Monte-Carlo algorithm\,\cite{Metropolis:1953am}. We underscore 
that the distribution of parameters is generated exclusively by the information 
that is presently known, namely, a single experimental measurement of the 
form factor that is encoded in the likelihood, and a set of theoretical biases 
that are embedded in the prior. The left-hand panel in Fig.\,\ref{Fig1} displays 
the correlation plot between the half-density radius $c$ and the surface 
diffuseness $a$ that define the SFermi distribution. As expected from just a 
single measurement, the model parameters are highly correlated. Indeed, the 
solid red line represents the functional relation implied by an ideal---{\sl i.e.,} 
error free and unbiased---single measurement of the form factor: 
$\FF{ch}(\qX{1};a,c)\!=\!0.210$. In turn, such a distribution of parameters 
generates the highly-asymmetric probability distribution function for the charge 
radius of ${}^{208}$Pb displayed on the right-hand panel of Fig.\,\ref{Fig1}. For 
comparison, the experimental value is 
$R_{\rm ch}^{208}\!=\!5.5012(13)\,{\rm fm}$\,\cite{Angeli:2013}.

\begin{figure}[ht]
 \begin{center}
  \includegraphics[width=0.45\linewidth]{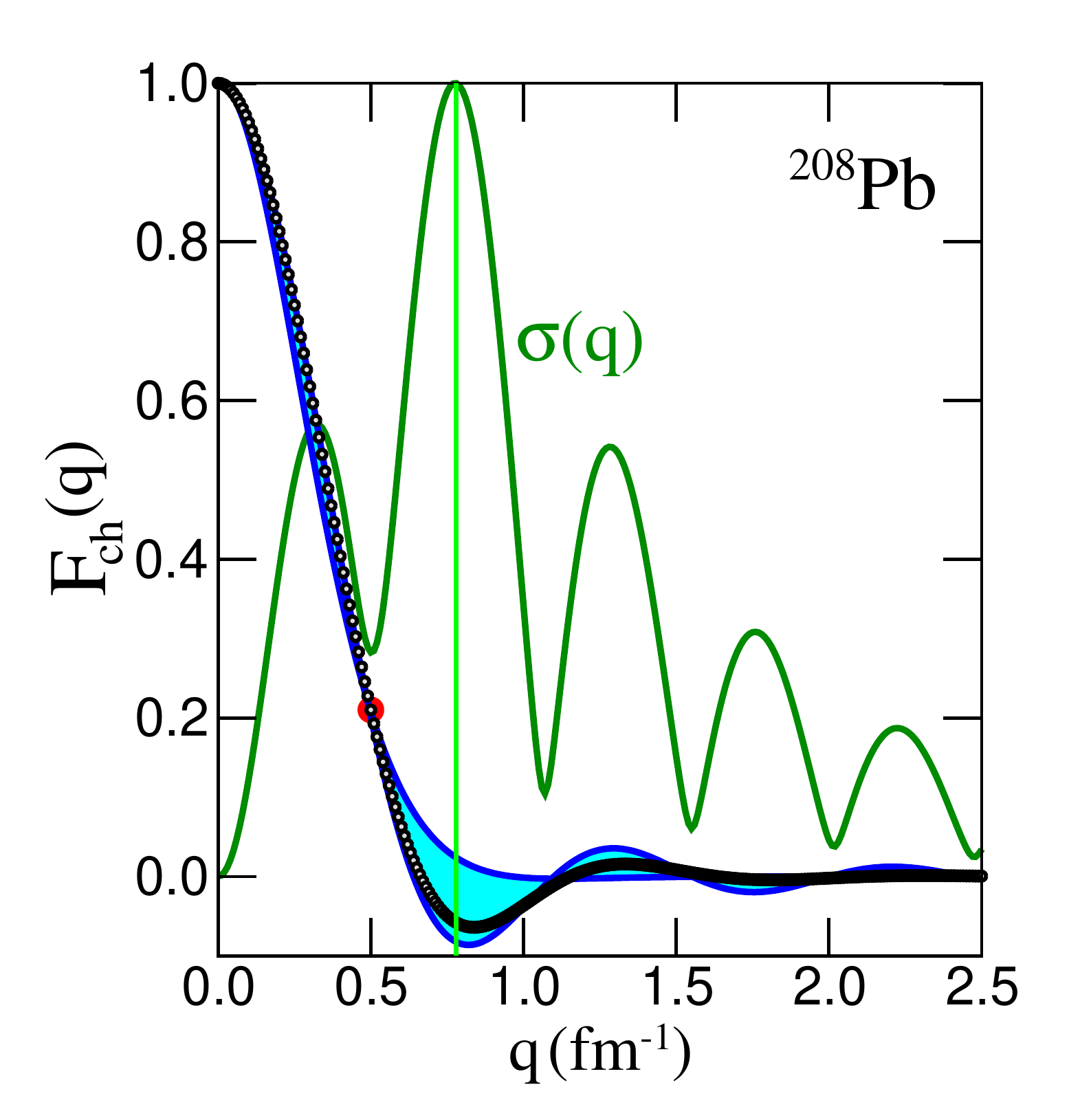} 
  \caption{(color online) Variability in the charge form factor of ${}^{208}$Pb 
  generated by the distribution of parameters displayed in Fig.\,\ref{Fig1}. 
  Also shown with the green solid line is the relative uncertainty in the model
  predictions as a function of the momentum transfer.} 
  \label{Fig2}
 \end{center} 
\end{figure}

As already alluded to and now clearly displayed in Fig.\,\ref{Fig2}, there is a very large 
(indeed infinite!) number of pairs $(c,a)$ that pass through the single experimental 
point (indicated by the red circle). However, the plot is highly informative because it
identifies the region of largest variability among the many models that satisfy the
experimental constraint. This variability is quantified in terms of the relative variance 
in the model predictions (green solid line) which is maximized in the region around
$\qX{2}\!=\!0.8\,{\rm fm}^{-1}$. This situation is highly favorable because the 
momentum transfer is large enough for the parity-violating asymmetry to be 
``sizable" (as the asymmetry scales with $q^{2}$) but not overly large for the 
nuclear form factor to be highly suppressed. 

%%%%%%%%%%%%%%%%%%%%%%%%%%%%%%%%%%%%%%%%%%%%%%%%%%%%%
%%%%%%%%%%%%%%%%%%%%%%%%%%%%%%%%%%%%%%%%%%%%%%%%%%%%%

Having identified the second $q$-point, we are now ready to answer the central question 
posed in this manuscript: \emph{How accurately can we describe the form factor of 
{\rm ${}^{208}$Pb} by measuring only two points?} To answer this question we repeat the 
same procedure that we have just implemented but now with a likelihood function augmented 
by a second value of the charge form factor at the newly identified momentum transfer of 
$\qX{2}\!=\!0.8\,{\rm fm}^{-1}$, namely, $\FF{ch}(\qX{2})\!=\!-0.0614$. For simplicity, we 
assume the same prior distribution and the same experimental error as before; that is, 
$\Delta F_{\rm exp}(\qX{1})\!=\!\Delta F_{\rm exp}(\qX{2})\!=\!0.005$. 

\begin{figure}[ht]
 \begin{center}
  \includegraphics[width=0.475\linewidth]{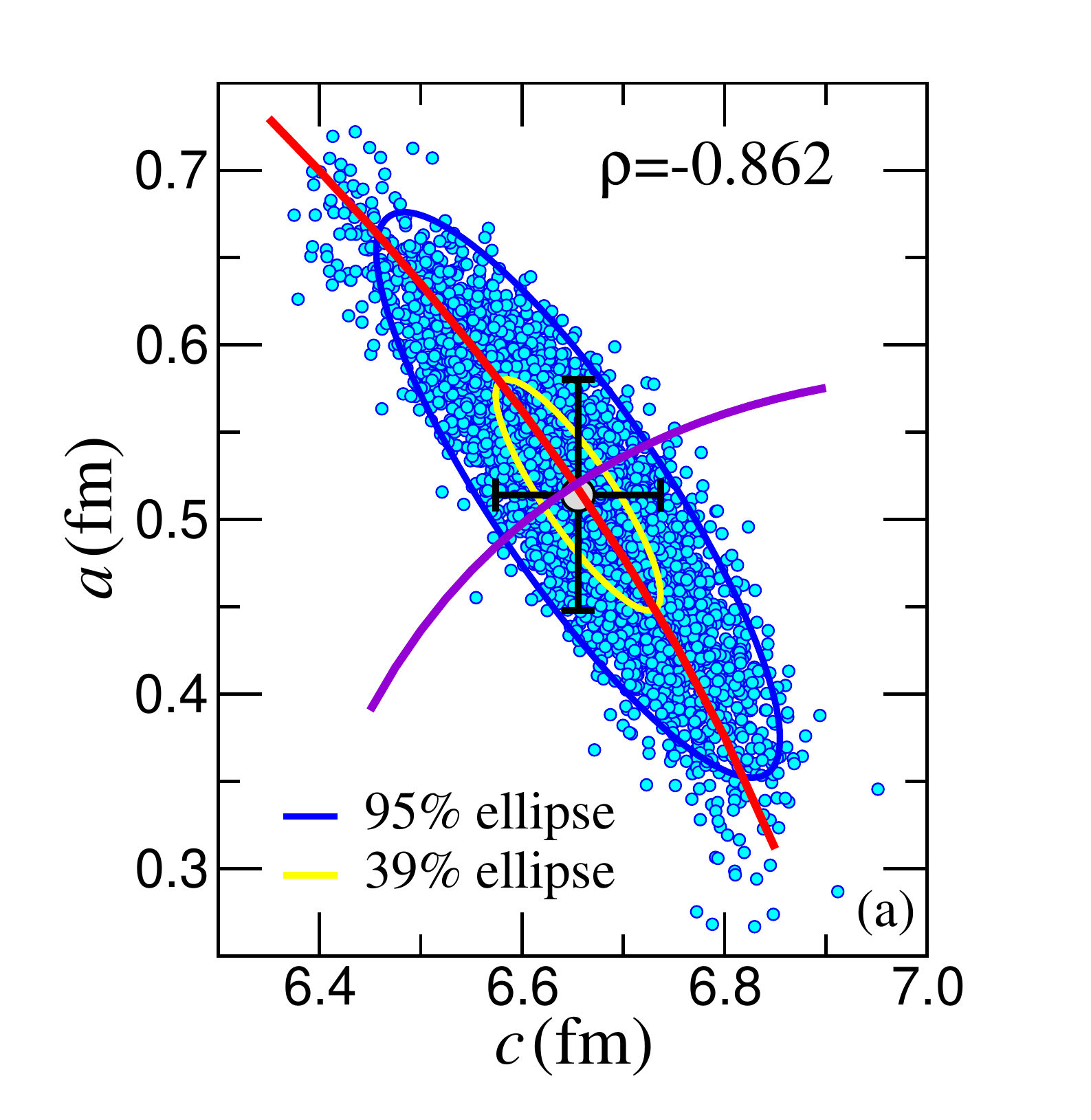} 
  \includegraphics[width=0.475\linewidth]{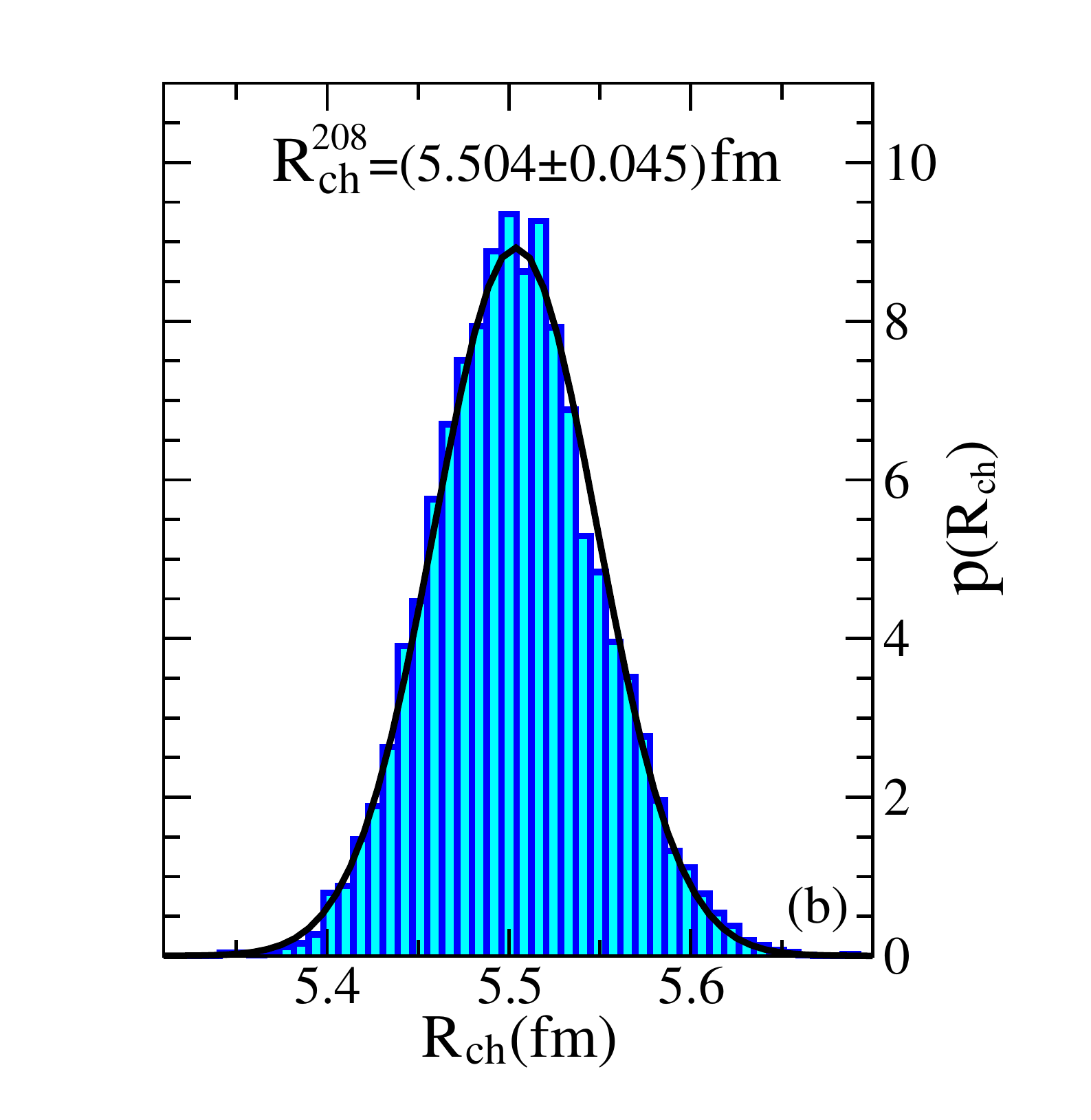}
  \caption{(color online) (a) Correlation plot between the half-density radius $c$ and the 
  surface diffuseness $a$ that define the symmetrized Fermi function. The number
  of points represent the raw data obtained from the Monte-Carlo simulation. Also shown 
  are the 39\% and 95\% confidence ellipses. The solid red and purple lines represents 
  the functional form obtained from solving the equations: $F_{\rm ch}(q_{1};a,c)\!=\!0.210$
   $F_{\rm ch}(q_{2};a,c)\!=\!-0.061$. (b) Probability distribution function for the charge 
   radius of ${}^{208}$Pb obtained from the Monte-Carlo simulation. The black solid line 
   represents a fit to a Gaussian probability distribution.}
  \label{Fig3}
 \end{center} 
\end{figure}

The improvement in our knowledge of the underlying symmetrized Fermi function is readily 
apparent in Fig.\,\ref{Fig3}. This figure is the analog of Fig.\,\ref{Fig1}, although note the 
disparity in scales between the two figures. The distribution of parameters is displayed 
on the left-hand panel of Fig.\,\ref{Fig3} together with the 39\% (in yellow) and 95\% (in blue) 
confidence ellipsoids. Note that now the posterior distribution is constrained by two functional 
relations: $\FF{ch}(\qX{1};a,c)\!=\!0.210$ (solid red line) and $\FF{ch}(\qX{2};a,c)\!=\!-0.061$ 
(solid purple line) which together provide nearly ``orthogonal'' constraints that set 
stringent limits on both the half-density radius $c$ and the surface diffuseness 
$a$; see Table\,\ref{Table2}. Finally, the right-hand panel in Fig.\,\ref{Fig3} shows 
the probability distribution function for the charge radius of ${}^{208}$Pb [see 
Eq.\,(\ref{SFMoments})]; along with the best Gaussian fit. Thus, the theoretical 
prediction for the charge radius of ${}^{208}$Pb obtained  from the knowledge of only two 
experimental points is $R_{\rm ch}^{208}\!=\!5.504(45)\,{\rm fm}$, which compares very 
favorably with the corresponding experimental value of 
$R_{\rm ch}^{208}\!=\!5.5012(13)\,{\rm fm}$\,\cite{Angeli:2013}.

%%%%%%%%%%%%%%%%%%%%%%%%%%%%%%%%%%%%%%%%%%%%%%%%%%%%%
%%%%%%%%%%%%%%%%%%%%%%%%%%%%%%%%%%%%%%%%%%%%%%%%%%%%%

%%%%%%%%%%%%%%%%%%%%%%%%%%%%%%%%%%%
\begin{table}[h]
  \begin{tabular}{|r|r|}
   \hline
    SFermi \hspace{0.55cm} & Helm\hspace{1cm}   \\
   \hline
    \rule{0pt}{9pt}  
    $c=6.655\pm0.081$ & $R_{0}=6.785\pm0.057$   \\
    $a=0.514\pm0.066$ & $\sigma=0.913\pm0.116$  \\
    $R_{\rm ch}=5.504\pm0.045$ & $R_{\rm ch}=5.492\pm0.041$ \\
   \hline
  \end{tabular}
  \caption{Average values and corresponding theoretical uncertainties 
  generated from the posterior distribution for a symmetrized Fermi and
  Helm form factors. Also shown are the predictions for the charge radius 
  of ${}^{208}$Pb, which should be compared against the experimental 
  value of $R_{\rm exp}^{208}\!=\!(5.5012\pm 0.0013)\,{\rm fm}$\,\cite{Angeli:2013}.
  All quantities are given in fm.}
  \label{Table2}
 \end{table}
%%%%%%%%%%%%%%%%%%%%%%%%%%%%%%%%%%%

\begin{figure}[ht]
 \begin{center}
  \includegraphics[width=0.475\linewidth]{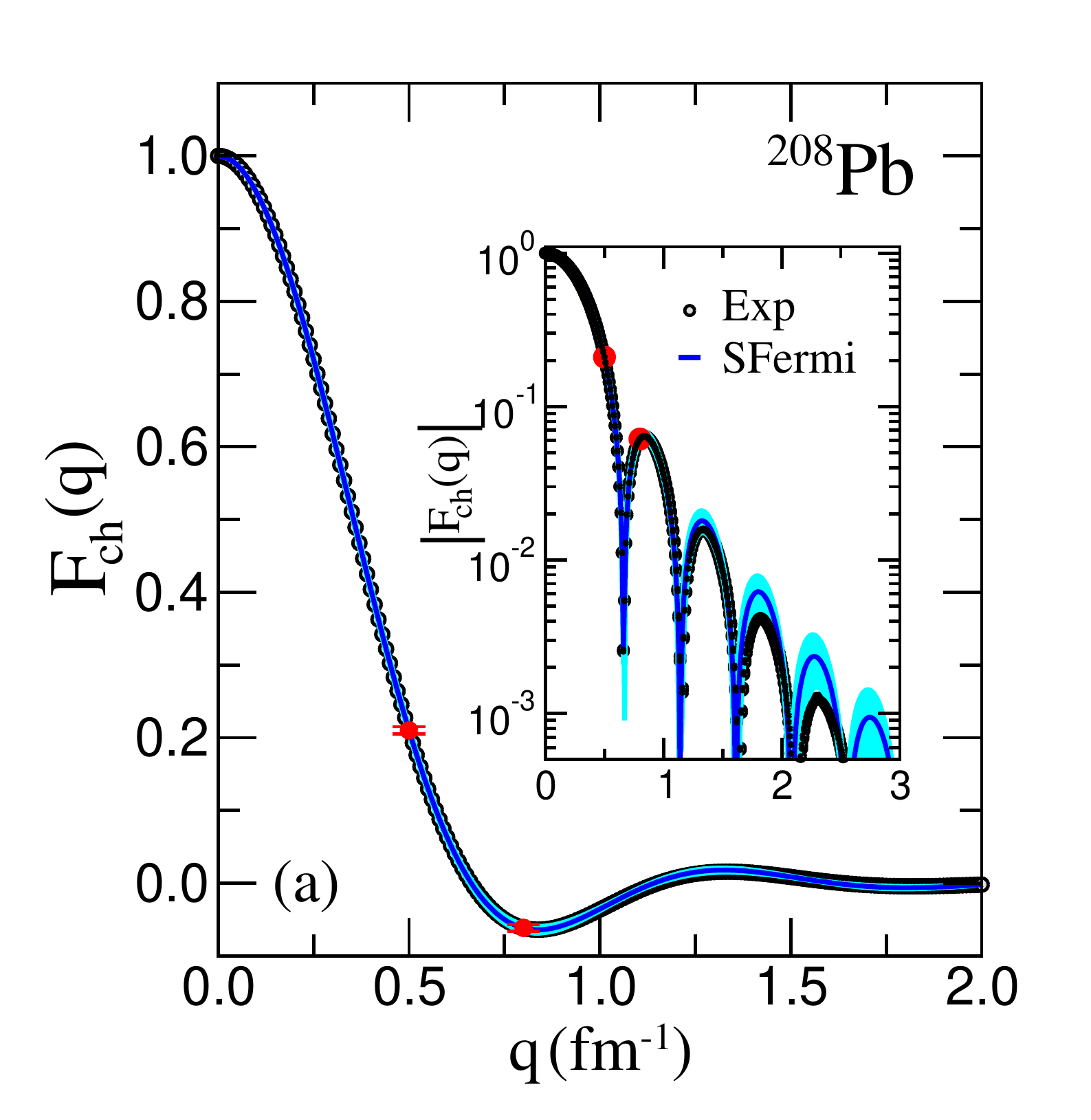} 
  \hspace{5pt}
  \includegraphics[width=0.475\linewidth]{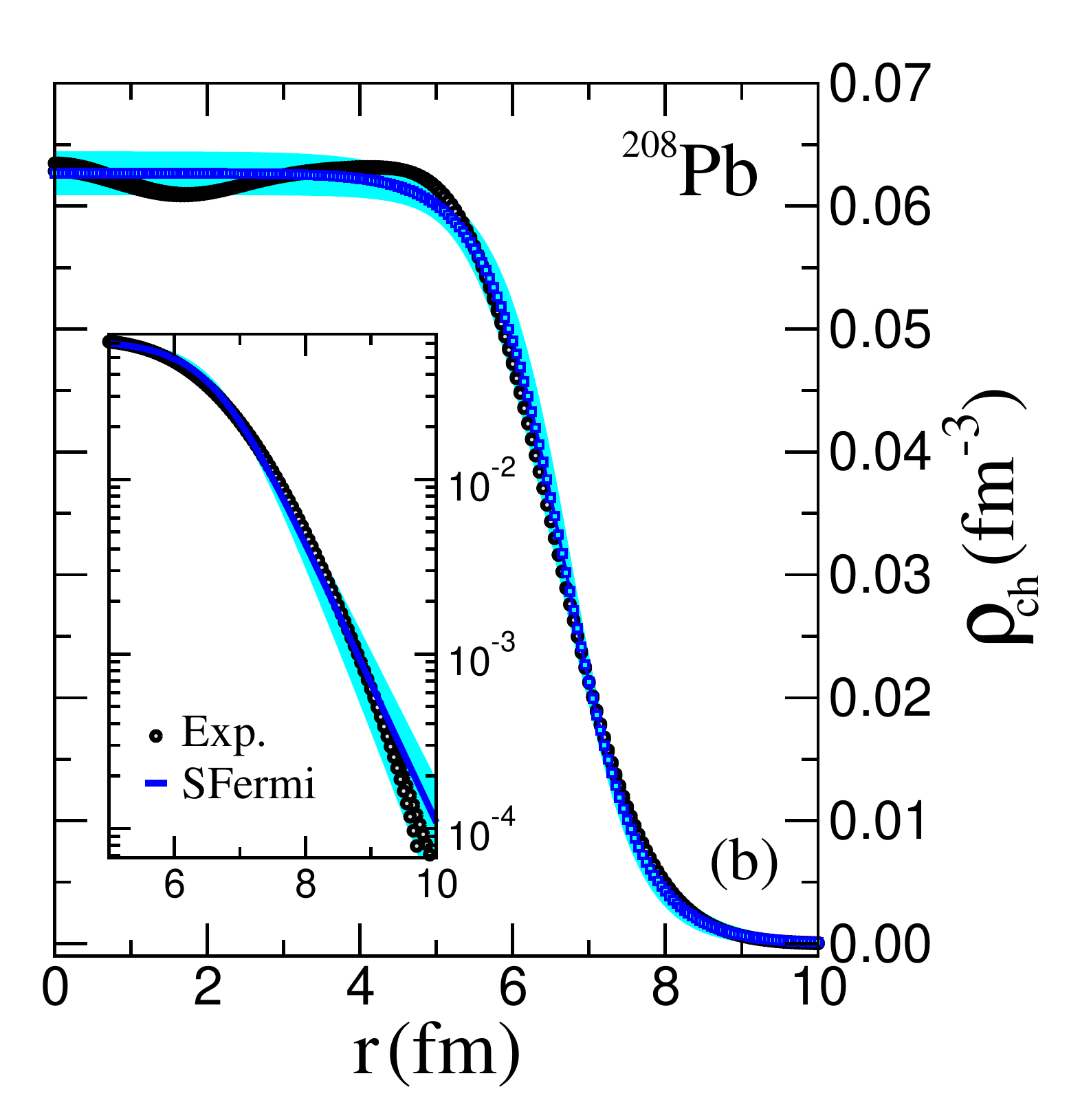}
  \caption{(color online) (a) Charge form factor and (b) the corresponding charge density 
  of ${}^{208}$Pb. The two red points on the left-hand panel represent the sole input used 
  in the calibration of the symmetrized Fermi function. The theoretical predictions are 
  displayed by an uncertainty band (in cyan) and the experimental data is from 
  Ref.\,\cite{DeJager:1987qc}.} 
  \label{Fig4}
 \end{center} 
\end{figure}

Having calibrated the parameters of the symmetrized Fermi function
using two experimental points and a fairly unconstrained prior 
distribution, we are now in a position to examine the overall agreement
between the theoretical predictions and the experimental data for the
\emph{entire} charge form factor. This is shown in Fig.\,\ref{Fig4}a using 
both linear and logarithmic scales. The two isolated red points represent
the two experimental measurements that were used to calibrate the model 
parameters ($c$ and $a$). In turn, the dense collection of black points 
represents the full experimental form factor\,\cite{DeJager:1987qc} 
that we aim to reproduce. Our theoretical predictions are displayed 
with a blue solid line together with the theoretical-uncertainty band 
shown in cyan. On a linear scale, it is difficult to discern the agreement
(or lack-thereof) between theory and experiment. Moreover, on a linear 
scale it is also difficult to appreciate the diffractive oscillations modulated 
by an exponential envelope that are the hallmark of the nuclear form 
factor. Thus, we display on the inset in Fig.\,\ref{Fig4}a the absolute 
value of the form factor using a logarithmic scale. The diffractive oscillations 
(controlled by $c$) and the exponential envelope (controlled by $a$) are now 
easily discernible. We observe a fairly good agreement between theory and
experiment over several diffractive maxima up to momentum transfers well 
beyond the value of the second point ($\qX{2}\!=\!0.8\,{\rm fm}^{-1}$). 
However, at the largest momentum transfers displayed in the figure, 
{\sl i.e.,} $q\!\gtrsim\!2.5\,{\rm fm}^{-1}$, there is a clear deterioration in 
the model predictions. Finally, the associated charge density of ${}^{208}$Pb 
is displayed in Fig.\,\ref{Fig4}b. Although it provides an excellent description 
of the experimental data at large distances as evinced in the inset, it fails to 
account for the experimental ``dip" in the nuclear interior, which correlates 
with the deterioration of the theoretical predictions at large momentum transfers. 
Note that in contrast, accurately-calibrated mean-field models tend to 
\emph{overestimate} the dip in the nuclear interior which is sensitive to
shell effects\,\cite{Fattoyev:2010mx}; see Fig.\,\ref{Fig6}.

%%%%%%%%%%%%%%%%%%%%%%%%%%%%%%%%%%%%%%%%%%%%%%%%%%%%%
%%%%%%%%%%%%%%%%%%%%%%%%%%%%%%%%%%%%%%%%%%%%%%%%%%%%%

\begin{figure}[ht]
 \begin{center}
  \includegraphics[width=0.475\linewidth]{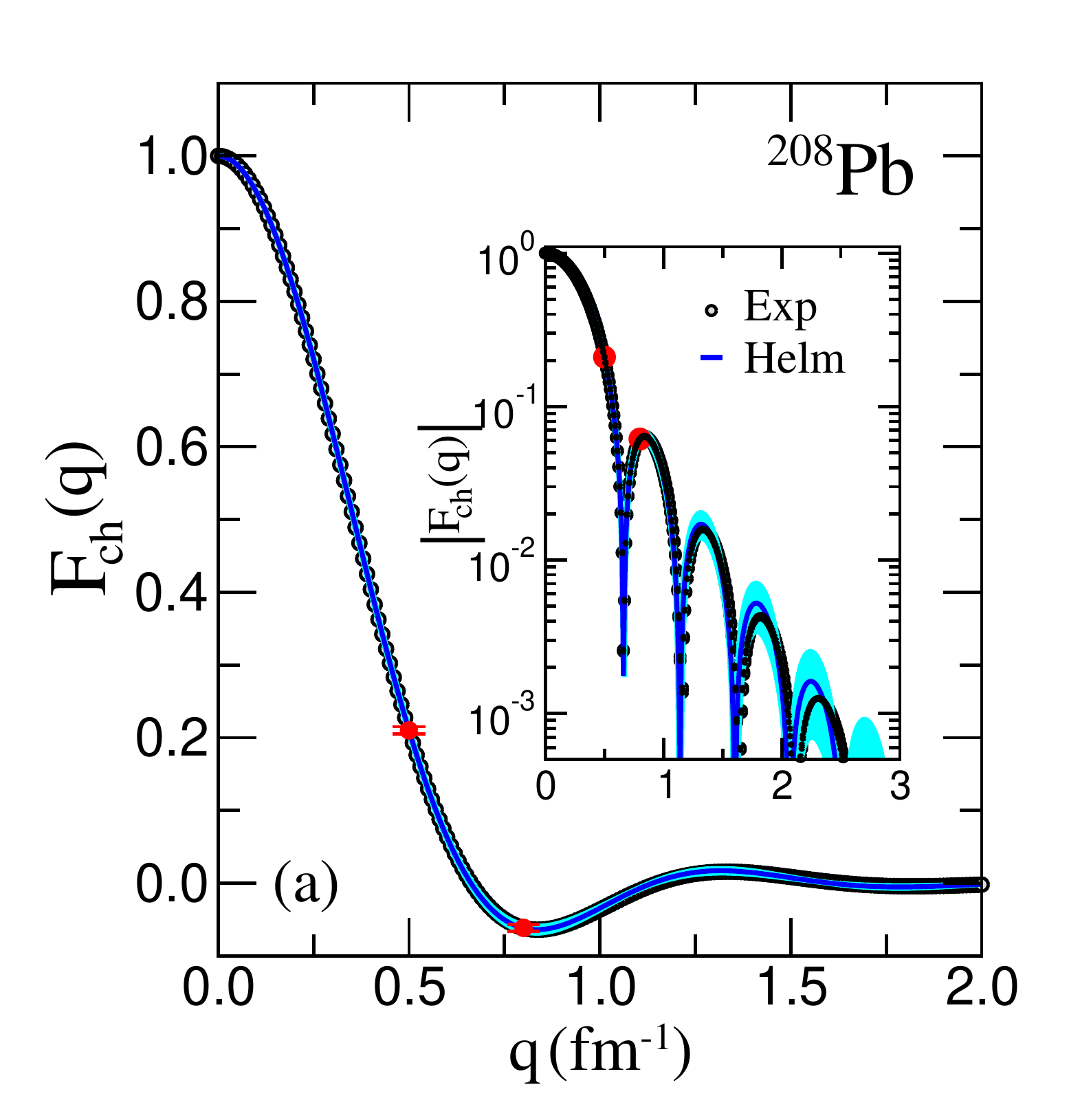} 
  \hspace{5pt}
  \includegraphics[width=0.475\linewidth]{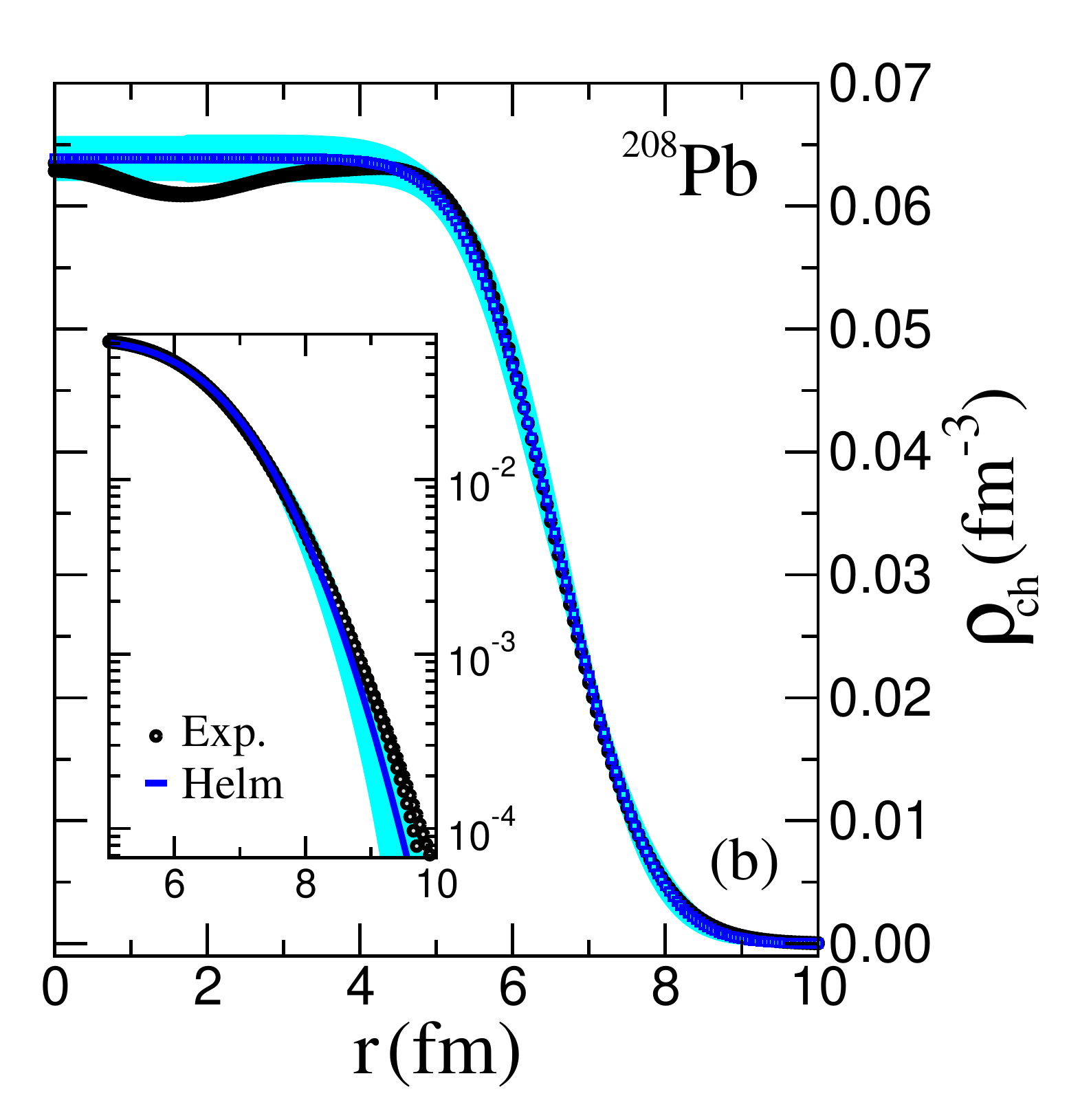}
  \caption{(color online) (a) Charge form factor and (b) the corresponding charge density 
  of ${}^{208}$Pb. The two red points represent the sole input used in the calibration 
  of the Helm  function. The theoretical predictions are displayed by an uncertainty band 
  (in cyan) and the experimental data is from Ref.\,\cite{DeJager:1987qc}.} 
  \label{Fig5}
 \end{center} 
\end{figure}

\begin{figure}[ht]
 \begin{center}
  \includegraphics[width=0.6\linewidth]{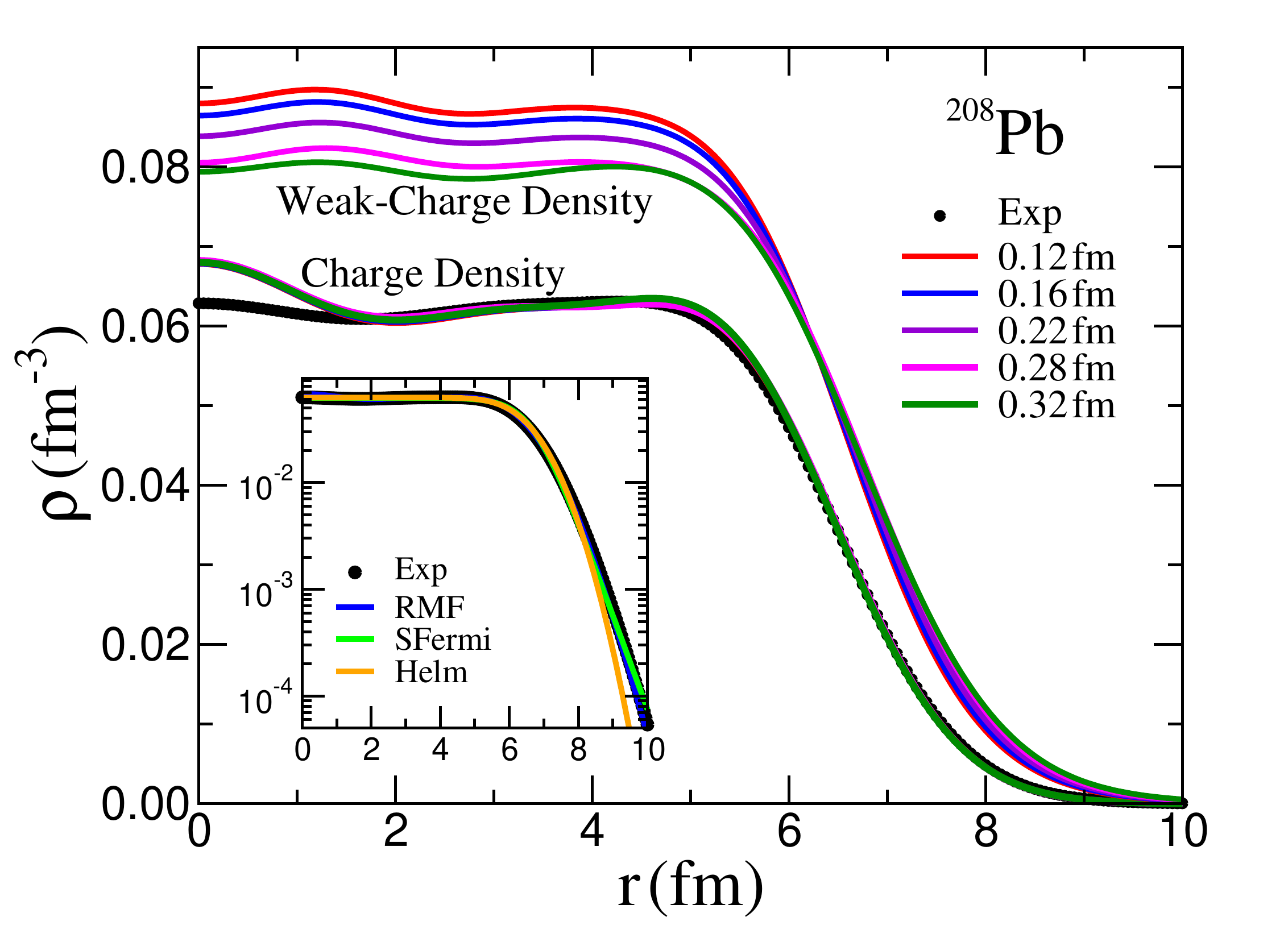} 
  \caption{(color online) Charge and weak-charge density of ${}^{208}$Pb
  as predicted by a collection of accurately calibrated mean-field models.
  The labels indicate the predicted neutron-skin thickness of ${}^{208}$Pb.
  The inset displays the fastest Gaussian falloff of the Helm form factor 
  relative to the exponential falloff of symmetrized Fermi function. The 
  experimental data is from Ref.\,\cite{DeJager:1987qc}.} 
  \label{Fig6}
 \end{center} 
\end{figure}

For completeness, we display in Fig.\,\ref{Fig5} the form factor and corresponding 
spatial density of ${}^{208}$Pb---but now using the Helm representation. Here too 
the agreement with experiment is fairly good and underscores the fact that any 
two-parameter function that properly encapsulates the diffractive oscillations an the 
exponential falloff of the form factor is likely to provide an adequate description of 
the data, at least at low momentum transfers. Naturally, the great virtue of the 
symmetrized Fermi and Helm parameterizations is that both the spatial density 
and the form factor are known in closed analytic form. However, a distinct advantage 
of the former over the latter is that it displays an \emph{exponential} rather than a
Gaussian falloff at large distances. 

%%%%%%%%%%%%%%%%%%%%%%%%%%%%%%%%%%%%%%%%%%%%%%%%%%%%%
%%%%%%%%%%%%%%%%%%%%%%%%%%%%%%%%%%%%%%%%%%%%%%%%%%%%%

As mentioned repeatedly earlier, the main goal of this manuscript is to assess using exclusively 
statistical methods and physical insights the impact of a second electroweak measurement 
of the weak form factor of ${}^{208}$Pb. In particular, we aim to quantify the experimental
precision required in the determination of the weak-charge (or neutron) radius of ${}^{208}$Pb 
to have a strong impact on both nuclear structure and astrophysics. Using the charge form 
factor of ${}^{208}$Pb as a proxy, we found that by measuring two suitable points with relative 
small, yet attainable, errors, the charge radius of ${}^{208}$Pb could be reproduced accurately 
with a precision of about $0.04\,{\rm fm}$. It is therefore natural to ask how meaningful would 
a measurement of the \emph{weak-charge} radius of ${}^{208}$Pb to this precision be on 
constraining the density dependence of the symmetry energy? 

To elucidate this question we display in Table\,\ref{Table3} predictions for several relevant 
quantities computed with a recent set of accurately calibrated relativistic mean field models.
These models are constrained by the same isoscalar sector but differ in a single isovector 
assumption, namely,  the choice of the neutron skin thickness of 
${}^{208}$Pb\,\cite{Chen:2014mza}. Although the set of models is relatively small, note that 
the theoretical spread in $R_{\rm wk}^{208}$ is nearly five times as large as the assumed 
$(0.04\,{\rm fm})$ experimental precision. Pictorially, the imprint of the isovector sector is 
illustrated in Fig.\,\ref{Fig6}. Indeed, whereas the charge density remains practically unchanged, 
significant differences emerge in the weak-charge density, as the latter is dominated by the 
neutron distribution. Shown in the inset on a logarithmic scale are symmetrized Fermi and Helm 
fits to the RMF charge density that evinced the more realistic exponential falloff of the 
SFermi density. As a figure of merit, we can establish that if $R_{\rm wk}^{208}$ is
relatively small, {\sl i.e.,} in the $R_{\rm wk}^{208}\!=\!(5.64\!-\!5.72)\,{\rm fm}$ range, 
or within the assumed $\pm0.04\,{\rm fm}$ uncertainty, one could 
constrain the slope of the symmetry energy to about $15\,{\rm MeV}$ and the radius of 
a $1.4\,M_{\odot}$ neutron star to within $1.2\,{\rm km}$. Of course, these estimates are
based on a very limited set of RMF models that suffer from their own limitations and
theoretical biases. Yet, our conclusions appear consistent with other studies that 
incorporate a very large ensemble of reasonable nuclear energy density 
functionals\,\cite{RocaMaza:2011pm,Piekarewicz:2012pp}.

%%%%%%%%%%%%%%%%%%%%%%%%%%%%%%%%%%%%%%%%%%%%%%%%%%%%%
\begin{widetext}
\begin{center}
\begin{table}[h]
\begin{tabular}{|l||c|c|c|c|c|c|}
 \hline
 \rule{0pt}{9pt}  
 Model & $R_{\rm ch}^{208}$ & $R_{\rm wk}^{208}$ & 
 $R_{\rm wk}^{208}\!-\!R_{\rm ch}^{208}$ & 
 $F_{\rm wk}^{\rm PREX}$ & $L$ & $R(1.4M_{\odot})$\\
 \hline
 \hline
 RMF-012 &  5.504 & 5.636 & 0.132 & 0.239  & 48.254 & 12.400\\
 RMF-016 &  5.499 & 5.667 & 0.168 & 0.234  & 50.961 & 12.839\\
 RMF-022 &  5.496 & 5.722 & 0.226 & 0.226 & 63.524 & 13.609\\
 RMF-028 &  5.495 & 5.790 & 0.295 & 0.216 & 112.644 & 14.234\\
 RMF-032 &  5.489 & 5.822 & 0.333 & 0.212 & 125.626 & 14.718\\
 \hline
\end{tabular}
\caption{Predictions from a set of accurately calibrated relativistic 
mean-field models\,\cite{Chen:2014mza} for the charge radius, weak-charge 
radius, and their difference for ${}^{208}$Pb (all in fm). Also shown is the
weak-charge form factor at the PREX momentum transfer, the 
slope of the symmetry energy $L$ (in MeV) and the radius of a
$1.4\,M_{\odot}$ neutron star (in km). Some of these quantities 
may be compared against the following experimental values: 
$R_{\rm ch}^{208}\!=\!5.5012(13)\,{\rm fm}$\,\cite{Angeli:2013}, 
$F_{\rm wk}^{\rm PREX}\!=\!0.204(28)$, and 
$R_{\rm wk}^{208}\!=\!5.826(181)\,{\rm fm}$\,\cite{Horowitz:2012tj}.} 
\label{Table3}
\end{table}
\end{center}
\end{widetext}
%%%%%%%%%%%%%%%%%%%%%%%%%%%%%%%%%%%%%%%%%%%%%%%%%%%%%

%%%%%%%%%%%%%%%%%%%%%%%%%%%%%%%%%%%%%%%%%%%%%%%%%%%%%
%%%%%%%%%%%%%%%%%%%%%%%%%%%%%%%%%%%%%%%%%%%%%%%%%%%%%

%%%%%  SECTION 4 - Conclusions  %%%%%
\section{Conclusions}
\label{conclusions}

Almost 80 years ago and shortly after the discovery of the neutron by Chadwick,
Bethe and Weizs\"acker described the atomic nucleus as a two-component quantum 
drop with a constant interior density and a nearly universal surface. Since then, elastic 
electron scattering experiments have provided a detailed map of the charge distribution 
that validates the simple picture of Bethe and Weizs\"acker. Indeed, to a large extent 
the nuclear charge density can be accurately described by only two parameters: a radius 
and a diffuseness. These two parameters leave their imprint in the characteristic diffractive 
oscillations modulated by an exponential falloff of the charge form factor. In this way, 
elastic electron scattering has provided the most accurate map of the distribution of 
charge in the nucleus.

Unfortunately, our picture of the corresponding \emph{weak-charge} density is fairly crude.
Whereas the charge density is dominated by the protons, the weak-charge density is dominated 
by the neutrons, as the weak neutral $Z^{0}$ boson couples preferentially to the neutrons. However, 
probing the neutron distribution is enormously challenging. Strongly interacting probes such as 
pions and protons couple strongly to neutrons but these reactions are plagued by hadronic 
uncertainties. Instead, parity-violating electron scattering is clean and model independent but the 
measured asymmetry is very small. Fortunately, in a pioneering measurement, the PREX 
collaboration used parity-violating electron scattering to extract the weak-charge form factor of 
${}^{208}$Pb at a single value of the momentum transfer\,\cite{Abrahamyan:2012gp}.  

In this manuscript we used standard statistical methods and physical insights to assess
the impact of a second electroweak measurement of the weak-charge form factor of ${}^{208}$Pb.
To do so, we introduced---or rather re-introduced\,\cite{Sprung:1997}---the two-parameter 
\emph{symmetrized Fermi function}, that is practically identical to the conventional Fermi function, 
but with far superior analytic properties. Indeed, the symmetrized Fermi function has a form factor 
that is known exactly in closed analytic form. By using such a parametrization, we estimated the 
accuracy and precision by which the root-mean-square radius of the distribution may be extracted 
from a single experimental measurement. Given that the symmetrized Fermi function---or any other
realistic parametrization---requires the determination of two parameters, it is perhaps not surprising 
that we found a large number of combinations of parameters that satisfy the single experimental 
constraint and, thus, a resulting RMS radius that was neither accurate nor precise. So the natural 
follow-up task involved estimating the potential improvement in the determination of the RMS radius
from a measurement of the form factor at a suitably chosen second point. To address this task we
used the exquisitely known experimental charge form factor of ${}^{208}$Pb as a proxy for the weak
charge form factor. To select the second point we examined the largest spread in the predictions of
all the symmetrized Fermi models that satisfy the original (one point) constraint. Based on the
largest variability, the optimal second point was identified near the first diffraction maximum 
({\sl i.e.,} near the first maximum in $|F(q)|$ away from $q\!=\!0$). Incorporating this second point
into the posterior probability density resulted in a significant improvement. First, we observed that 
the two measurements provide nearly ``orthogonal'' constraints that lift the original degeneracy among 
the parameters. Second, we obtained a RMS radius that is both accurate 
($R_{\rm ch}^{208}\!=\!5.504\,{\rm fm}$) and precise ($\Delta R_{\rm ch}^{208}\!=\!0.045\,{\rm fm}$)
as compared with the enormously precise experimental value of 
$R_{\rm exp}^{208}\!=\!5.5012(13)\,{\rm fm}$. Finally, when compared against the entire experimental 
charge form factor, the two-parameter symmetrized Fermi function provides an excellent description 
over several diffraction maxima. 
 
So how accurately could one constrain the density dependence of the symmetry energy if
the weak-charge radius of ${}^{208}$Pb could be measured with a precision of about $0.04\,{\rm fm}$?
Based on a limited set of accurately calibrated RMF models, we estimated that a $\pm0.04\,{\rm fm}$ 
determination of $R_{\rm wk}^{208}$ would translate into an overall constraint on the slope of the 
symmetry energy of about $15\,{\rm MeV}$. That is, if the symmetry energy is soft leading to a
thin neutron skin, then $L$ was predicted to lie in the $L\!\approx\!(48\!-\!63)\,{\rm MeV}$
range. Although these RMF models are hindered by their own limitations and theoretical biases,
more comprehensive studies using a large set of both non-relativistic and relativistic energy density 
functionals are consistent with these results. In fact, they often suggest even more stringent 
limits! While we recognize that achieving a $\pm0.04\,{\rm fm}$ (or better) precision represents 
an enormous experimental challenge, the strong impact of a second measurement of the weak form
factor of ${}^{208}$Pb on both nuclear and neutron-star structure may be worth the significant effort.

\vfill\eject

\begin{acknowledgments}
This material is based upon work supported by the U.S. Department of Energy 
Office of Science, Office of Nuclear Physics under Award Number 
DE-FD05-92ER40750.
\end{acknowledgments}

%\bibliography{../ReferencesJP}
\bibliography{LeDEx.bbl}

\end{document}